\documentclass[a4paper]{article}

\usepackage[colorlinks,
            linkcolor=blue,
            anchorcolor=blue,
            citecolor=blue,
            ]{hyperref}

\def\bt{\begin{theorem}}
\def\et{\end{theorem}}
\def\bp{\begin{proposition}}
\def\ep{\end{proposition}}
\def\bc{\begin{corollary}}
\def\ec{\end{corollary}}
\def\bo{\begin{proof}}
\def\eo{\end{proof}}
\def\bx{\begin{example}}
\def\ex{\end{example}}
\def\br{\begin{remark}}
\def\er{\end{remark}}
\def\bl{\begin{lemma}}
\def\el{\end{lemma}}

\def\bn{\begin{definition}}
\def\en{\end{definition}}
\def\ba{\begin{array}}
\def\ea{\end{array}}
\def\be{\begin{equation}}
\def\ee{\end{equation}}
\def\bd{\begin{description}}
\def\ed{\end{description}}
\def\bu{\begin{enumerate}}
\def\eu{\end{enumerate}}
\def\bi{\begin{itemize}}
\def\ei{\end{itemize}}

\usepackage{amsmath,amssymb,graphicx}
\usepackage{subfigure}
\usepackage{epsfig}
\usepackage{epstopdf}
\usepackage{lscape}
\usepackage{mathrsfs}

\usepackage{algorithmic}
\usepackage[ruled]{algorithm}

\usepackage{multirow}
\usepackage{verbatim}

\usepackage{color}

\newtheorem{lemma}{Lemma}
\newtheorem{theorem}{Theorem}
\newtheorem{proposition}{Proposition}
\newtheorem{corollary}{Corollary}
\newtheorem{remark}{Remark}

\def\i{{\bf i}}

\usepackage{longtable}

\def\ds{\displaystyle}

\title{ Quantum Regularized Least Squares Solver with Parameter Estimate }

\author{Changpeng Shao\thanks{Academy of Mathematics and Systems Science, Chinese Academy of Sciences, Beijing 100190, P. R. China. (cpshao@amss.ac.cn). }
~and
Hua Xiang\thanks{School of Mathematics and Statistics, Wuhan University, Wuhan 430072, P. R. China. The work of the author was supported by the Natural Science Foundation of China under grants 11571265 and NSFC-RGC
No. 11661161017.  (hxiang@whu.edu.cn). }
}

\begin{document}
\maketitle

\begin{abstract}
In this paper we propose a quantum algorithm to determine the Tikhonov regularization parameter and solve the 
ill-conditioned linear equations. 
For regularized least squares problem with a fixed regularization parameter, we use the HHL algorithm and work on an extended matrix with smaller condition number. For the determination of the regularization parameter, we combine the classical L-curve and GCV function, and design quantum algorithms to compute the norms of regularized solution and the corresponding residual in parallel and locate the best regularization parameter by Grover's search. The quantum algorithm can achieve
a quadratic speedup in the number of regularization parameters and an exponential speedup
in the dimension of problem size.
\end{abstract}

{\bf Keywords}. HHL algorithm, Grover's search, Tikhonov regularization, TSVD,  Regularization parameter, L-curve, GCV.



\section{Introduction}

We consider an {\em ill-conditioned} linear system $Ax=b$ that arises from the discretization of some linear inverse
problem~
or of the linearized system of some nonlinear case \cite{Banks89,Engl96,Hansen_book2010},
where the coefficient matrix $A$ is of size $m\times n$   and the right hand side $b$ is obtained from measurement data. For the finite element discretization, $A$ is also a large sparse matrix.
For such ill-conditioned problem, the condition number $\kappa(A) = \|A^+\|\cdot \|A\|$ is large, where $A^+$ is the Moore-Penrose inverse. Thus some {\em regularization} techniques are needed in order to
achieve a meaningful solution, since the solution is very sensitive and can
be easily contaminated by the perturbation in the measurement data.     Tikhonov regularization is
one of the most popular and effective techniques, which converts the original linear system into the following regularized least squares problem (LSP)
\begin{equation}\label{x_regu}
\min_{x} \{\|Ax-b\|^2 + \mu^2 \|  x \|^2 \},
\end{equation}
where
constant $\mu$ is the so-called {\em regularization parameter} \cite{TikhonovArsenin_Wiley1977}.
By introducing the regularization parameter $\mu$, one can make a comprise between the sensitivity of the problem and the perturbation
of the measured data and thus greatly reduce the effect caused by the contamination of the noise in
the data.
In statistics the problem \eqref{x_regu} is the well-known ridge regression problem, and $\mu$ is also called the {\em ridge parameter}.
The Tikhonov regularization can be of the following more general form \cite{XiangZou_InverseProb15}
\begin{equation*}
\min_{x} \|A x - b\|^2 + \mu^2 \|L x\|^2,
\end{equation*}
where the matrix $L$ 
arises from the
discrete approximation to some differential operator, for  example,
the discrete Laplacian or gradient operator.

The LSP \eqref{x_regu}  can be solved by the singular value decomposition (SVD).
Suppose that we have the SVD of
matrix $A \in \mathbb{R}^{m \times n} (m \geq n)$, and it reads  $A = U \Sigma V^\dag$, where $U = (u_1, \cdots, u_m)$, $V = (v_1,
\cdots, v_n)$ are orthonormal matrices, and $\Sigma =
\text{diag}(\sigma_1, \cdots, \sigma_n) \in \mathbb{R}^{m \times n}$
with $\sigma_i~ (i=1,2,\cdots, n)$ being the singular
values.
Then the solution of (\ref{x_regu}), i.e., the Tikhonov regularized
solution $x_\mu$, can be expressed as
\begin{equation}\label{x_regu3}
x_{\mu} = \sum_{i=1}^n f_i \frac{u_i^\dag b}{\sigma_i} v_i ,
\end{equation}
 where $f_i = \sigma_i^2 /
(\sigma_i^2 + \mu^2)$ is the {\em Tikhonov  filter factor} \cite{Hansen_book2010}. 
For the case where the matrix $A$ arises from the discretization of some compact operator,
it has singular values of quite small magnitude. One can clearly see the necessity for introducing the regularization term in \eqref{x_regu}.

Another closely related popular regularization method is the truncated SVD method (TSVD), where just the largest $k$ singular values are kept while the other small ones are neglected. Using the resulting best low rank approximation of $A$, the TSVD regularized solution $x_k$ is given by
\be \label{x_regu-TSVD}
x_{k} = \sum_{i=1}^k   \frac{u_i^\dag b}{\sigma_i} v_i ,
\ee
where $k$ acts as the truncation parameter and is chosen such that the
noise-dominated small singular values are discarded.
From another viewpoint, \eqref{x_regu3} can be reduced to the TSVD solution \eqref{x_regu-TSVD} by replacing the filter factors $f_i$ in \eqref{x_regu3} by $0$'s and $1$'s
appropriately.

The success of the regularized solution highly depends on the choice of regularization parameter $\mu$ or $k$, which is our focus in this paper.
There are several popular techniques in the
literature to determine effective regularization parameters.
When the noise level is unknown, we may use some heuristic methods, such as the L-curve method \cite{Hansen_SIREV92,HansenLeary_SISC93}, the generalized cross-validation (GCV) function \cite{GolubHeathWahba_Technometrics79}, the quasi-optimality criterion, 
and so on.
When the noise level is known, the discrepancy principle, 
the monotone error rule, 
and the balancing principle 
can be applied (See \cite{XiangZou_InverseProb13} and the references therein).
In the following, we introduce the amplitude estimation in Section 2, which will be used for norm estimate in the choice of regularization parameter.
We consider the quantum regularized LS with a fixed parameter in Section 3, which is a preparation for the algorithm with variable parameters.
Finally we focus on the regularization parameter choice in Section 4. Classically, we choose a range of parameters $\mu$. For each parameter $\mu$, we solve the problem \eqref{x_regu} and obtain the corresponding solution $x_\mu$. From such a series of solutions $x_\mu$, we take a proper strategy to locate the possible best regularization parameter. The regularization parameter estimate is critical and very time consuming.
In this section we propose a quantum parallel procedure to speed up the parameter estimate.

\section{Amplitude estimation and its generalization}
\label{pst}

An technique we will apply to find the best regularization parameter is the amplitude estimation,
which was proposed in \cite{brassard-AA} as a byproduct of quantum phase estimation (QPE) and Grover's algorithm.
In the following, we first briefly review this quantum algorithm.
Then, we extend it into a more general form that can estimate amplitudes in parallel.

Let
\be\label{ini-form}
|\phi\rangle =U |0\rangle^{\otimes k} = \cos \theta|0\rangle|u\rangle + \sin \theta|1\rangle|v\rangle
\ee
be a quantum state, where $U$ is a unitary operator that can be implemented in time $O(T)$.
In the following we will show how to estimate $\theta$ in a quantum computer to accuracy $\epsilon$ with high success probability at least $1-\delta$.

Let $Z$ be the 2-dimensional Pauli-Z matrix that maps $|0\rangle$ to $|0\rangle$ and $|1\rangle$ to $-|1\rangle$.
Denote
$$
G=( 2|\phi\rangle\langle\phi| - I )(Z\otimes I)=U( 2|0\rangle^{\otimes k}\langle0|^{\otimes k} - I )U^\dag(Z\otimes I),
$$
which is similar to the rotation used in Grover's algorithm. We can check that
\[
G=\left(
   \begin{array}{rr} \vspace{.2cm}
     \cos2\theta  &~~ - \sin2\theta \\
     \sin2\theta  &~~   \cos2\theta \\
   \end{array}
 \right)
\]
in the space spanned by $\{|0\rangle|u\rangle,|1\rangle|v\rangle\}$. The eigenvalues of $G$ are $e^{\mp \i2\theta}$ and the corresponding eigenvectors are
$
|w_\pm\rangle=\frac{1}{\sqrt{2}}(|0\rangle|u\rangle\pm\i|1\rangle|v\rangle).
$

To apply QPE, we choose the initial state as $|0\rangle^{\otimes n}|\phi\rangle$, where $n=O(\log1/\delta\epsilon)$.
It can be rewritten as
$$
|0\rangle^{\otimes n}|\phi\rangle= \frac{1}{\sqrt{2}} |0\rangle^{\otimes n}(e^{\i\theta}|w_-\rangle + e^{-\i\theta}|w_+\rangle ).
$$
By QPE \cite{nielsen}, we obtain the following state
\be\label{final-form}
\frac{1}{\sqrt{2}} (e^{\i\theta}|y\rangle|w_-\rangle + e^{-\i\theta}|-y\rangle|w_+\rangle),
\ee
in time $O(T/\epsilon\delta)$, where $y\in \mathbb{Z}_{2^n}$ satisfies $|\theta-y\pi/2^n|\leq \epsilon$.
Here we ignored the circuit complexity to implement the Hadamard transformation
and the quantum Fourier transform in QPE, which equals $O(n^2) = O((\log1/\delta\epsilon)^2)$.
Performing a measurement on (\ref{final-form}), we get an $\epsilon$-approximate of $\theta$ or $-\theta$ with probability at least $1-\delta$.
From the approximation of $\pm\theta$, we can estimate the probabilities $|\sin\theta|^2$ and $|\cos\theta|^2$ efficiently.
Generally, $\delta$ is chosen as a small constant and so can be ignored in the complexity analysis.
The above is the main idea of amplitude estimation.

To further apply the information of $\theta$ to solve other problems,
such as the finding of the best regularization parameter studied in this paper,
instead of performing measurement at (\ref{final-form}), the following quantum state
is more useful
\be\label{final-form-1}
|0\rangle^{\otimes n}|\phi\rangle | f(\cos y\pi/2^n)\rangle,
\ee
where $f$ is a function defined in $\mathbb{C}$ such that $U_f:|x,y\rangle\mapsto|x,y\oplus f(x)\rangle$ is efficiently implemented.
The quantum state (\ref{final-form-1}) is obtained by adding a register at (\ref{final-form}) to store $f(\cos y\pi/2^n)$
and undoing the QPE. If $f$ is an elementary function, such as polynomial or cosine function,
then $U_f$ can be efficiently implemented in a quantum computer \cite{rieffel}.
All the functions we encountered in this paper are elementary, so in the following, we always
assume that $U_f$ is available.
For simplicity, we ignore the term $|0\rangle^{\otimes n}$ in (\ref{final-form-1}). Therefore,
we have

\bp \label{prop-AA}
Let $U$ be a unitary operator such that
$|\phi\rangle=U |0\rangle^{\otimes k} = \cos \theta|0\rangle|u\rangle + \sin \theta|1\rangle|v\rangle$.
Assume that $U$ can be implemented in time $O(T)$ in quantum computer.
Let $f$ be a complex function, then the following unitary transformation
\be \label{AA:general procedure}
|\phi\rangle |0\rangle \mapsto |\phi\rangle|f(\cos\tilde{\theta})\rangle
\ee
can be achieved in time $O(T/\epsilon)$, where $|\theta-\tilde{\theta}|\leq \epsilon$.
\ep

The unitary procedure (\ref{AA:general procedure}) can be implemented in parallel due to quantum superposition and parallelism
as the following corollary states.

\bc \label{coro-AA}
Let $U_1,\ldots,U_p$ be $p$ unitary operators that can be prepared in time $O(T)$ in quantum computer.
Assume that $|\phi_j\rangle = U_j |0\rangle^{\otimes k} = \cos \theta_j|0\rangle|u\rangle + \sin \theta_j|1\rangle|v\rangle$ for all $j$.
Let $\sum_{j=1}^p \alpha_j |j\rangle$ be a given quantum state
and $f_1,\ldots,f_p$ be $p$ complex functions.
Then the following unitary transformation
\be \label{AA:parallel procedure}
\sum_{j=1}^p \alpha_j |j\rangle |\phi_j\rangle |0\rangle \mapsto \sum_{j=1}^p \alpha_j |j\rangle |\phi_j\rangle |f_j(\cos\tilde{\theta}_j)\rangle
\ee
can be achieved in time $O(T/\epsilon)$, where $|\theta_j-\tilde{\theta}_j|\leq \epsilon$.
\ec

Combining Corollary \ref{coro-AA} and quantum minimum finding algorithm \cite{durr},
if $\alpha_j=1/\sqrt{p}$ for all $j$ in (\ref{AA:parallel procedure}), then we can find $j_0 = \arg \min_j f_j(\cos\theta_{j})$ in time $O(\sqrt{p}T/\epsilon)$.
The algorithm is a direct modification of \cite{durr}.

{\bf Quantum minimum finding algorithm  \cite{durr}:}
\begin{enumerate}
  \item Choose threshold index $1\leq y\leq p$ by performing a measurement at (\ref{AA:parallel procedure}).
  \item Repeat the following and interrupt it when the total running time is more than $22.5\sqrt{p}+1.4(\log p)^2$.
  \begin{enumerate}
    \item Initialize the memory as $\frac{1}{\sqrt{p}} \sum_{j=1}^p |j\rangle |\phi_j\rangle |f_j(\cos\tilde{\theta}_j)\rangle$.
          Mark every item $j$ for which $f_j(\cos\tilde{\theta}_j)<f_y(\cos\tilde{\theta}_y)$.
    \item Apply the amplitude amplification to improve the probability of marked items.
    \item Observe the first register: let $y'$ be the outcome.  If $f_{y'}(\cos\tilde{\theta}_{y'})<f_y(\cos\tilde{\theta}_y)$,
          then set $y'$ to be the new threshold index $y$.
  \end{enumerate}
  \item Return the index $y$.
\end{enumerate}


\bc \label{coro-AA-appliation}
Let $U_1,\ldots,U_p$ be $p$ unitary operators that can be prepared in time $O(T)$ in quantum computer, and let $f_1,\ldots,f_p$ be $p$ complex functions.
Assume that $|\phi_j\rangle = U_j |0\rangle^{\otimes k} = \cos \theta_j|0\rangle|u\rangle + \sin \theta_j|1\rangle|v\rangle$ for all $j$, and $|\theta_j-\tilde{\theta}_j|\leq \epsilon$.
Then we can find $j_0 = \arg \min_j f_j(\cos\theta_{j})$ in time $O(\sqrt{p}T/\epsilon)$.
\ec

\section{Quantum regularized LS Algorithm}

For the LSP \eqref{x_regu}, we need a solver for a system of linear equations.
There exist several methods that can be used to solve linear system in quantum computer, such as HHL algorithm \cite{HarrowHassidimLloyd_PRL09}, SVE \cite{wossnig}, the blocked-encoding framework \cite{chakraborty} and some extensions.
All these methods are affected by the condition number of the linear systems.
However, solving linear system is not the main objective of this paper, and we only focus on the HHL algorithm to solve linear systems, even though some other approaches have better dependence on the condition number and show better efficiency.
The HHL algorithm, the first quantum linear solver proposed by Harrow, Hassidim and Lloyd \cite{HarrowHassidimLloyd_PRL09} in 2009,  solves the linear system $Ax=b$, or equivalently
$\left(
   \begin{array}{cc}
     0 & A \\
     A^\dag & 0 \\
   \end{array}
 \right)\left(
   \begin{array}{cc}
     0 \\
     x \\
   \end{array}
 \right)=\left(
   \begin{array}{cc}
     b \\
     0 \\
   \end{array}
 \right)$.
This algorithm outputs a quantum state $| x \rangle$ proportional to $ \sum_{k=1}^n x_k |k\rangle$.
We can get the expectation value of a certain operator $M$ associated with $x$, i.e., $x^\dag M x$, by swap test \cite{buhrman}.
Similarly, given another quantum state $|c\rangle = \sum_{k=1}^n c_k |k\rangle$,
one can obtain an estimate of $\sum_{k=1}^n c_k x_k$ efficiently.

Actually the HHL algorithm finds the least squares solution of the problem $\min_x \|Ax-b\|$. That is, the HHL algorithm solves the linear system $A^\dag Ax=A^\dag b$  instead of the original one $Ax=b$.
So the HHL algorithm can be directly applied to the LSP.
The quantum algorithm  given in \cite{WiebeBraunLloyd_PRL12} seems to be the first application of HHL algorithm for the LSP.
Other quantum algorithms to solve linear regression can be found in \cite{chakraborty,kerenidis,SchuldSinayskiyPetruccione_PRA16, Wang_PRA17}.
The ordinary LSP is a special case of regularized
LSP \eqref{x_regu} with the regularization parameter $\mu=0$.
Solving the regularized LSP with a fixed regularization parameter $\mu$ has been considered in \cite{LiuZhang_TCS17}, based on the SVD of $A$.

It is easy to verify that the LSP \eqref{x_regu} is equivalent to the following form
\begin{equation}\label{x_regu1}
\min_{x} \left\|\left(
                  \begin{array}{c}
                    A \\
                    \mu I \\
                  \end{array}
                \right)x-\left(
                  \begin{array}{c}
                    b \\
                    0 \\
                  \end{array}
                \right) \right\|^2 .
\end{equation}
Our LSP solver starts from the SVD of the extended matrix $A_\mu:=\left(
                  \begin{array}{c}
                    A \\
                    \mu I \\
                  \end{array}
                \right)$.
Denote the condition numbers of $A$ and $A_\mu$ as $\kappa$ and $\kappa_\mu$, respectively.
Using the SVD of $A$, we can easily derive that the eigenvalues of $A_{\mu}^\dag A_{\mu}$ are $\sigma_i^2 + \mu^2~(i=1,\cdots, n)$, so the singular values of $A_\mu$ are $\sqrt{\sigma_i^2 + \mu^2}~(i=1,\cdots, n)$.
Let $\sigma_{\max}$ be the largest singular value and $\sigma_{\min}$ the smallest nonzero singular value of $A$, respectively.
If $A$ is of full column rank, then $\sigma_n = \sigma_{\min} \neq 0$, and
\be \label{relation:kappa}
\kappa_\mu = \sqrt{\frac{ \sigma_{\max}^2+\mu^2}{  \sigma_n^2+\mu^2}}
= \sqrt{\frac{\kappa^2 + (\mu/ \sigma_{\min})^2}{1+(\mu/ \sigma_{\min})^2}}. 
\ee
If $A$ is rank deficient or $m<n$, then $\sigma_n = 0$, and
\be \label{relation:kappa2}
\kappa_\mu = \sqrt{ \frac{\sigma_{\max}^2+\mu^2}{\mu^2} }= \sqrt{ \kappa^2 
\frac{\sigma_{\min}^2}{\mu^2} + 1 } .
\ee
For the ill-conditioned case where $\kappa \gg 1$, the regularization parameter $\mu$ is chosen such that $\mu \gg \sigma_{\min}$.
Then generally we have $\kappa_\mu \ll \kappa$.
So we may have a much smaller condition number for solving the LSP \eqref{x_regu} based on $A_\mu$, rather than that based on $A$.

The minimization problem \eqref{x_regu1} can be solved by applying the HHL algorithm to the following linear system
\be\label{linear-system-regu}
\left(
   \begin{array}{cc}
     0 & A_\mu \\
     A_\mu^\dag & 0 \\
   \end{array}
 \right)\left(
   \begin{array}{cc}
     0 \\
     x \\
   \end{array}
 \right)=\left(
   \begin{array}{cc}
     b \\
     0 \\
   \end{array}
 \right).
\ee
Denote $\widetilde{A}_\mu=\left(
   \begin{array}{cc}
     0 & A_\mu \\
     A_\mu^\dag & 0 \\
   \end{array}
 \right)$ and $\tilde{b}=\left(
   \begin{array}{cc}
     b \\
     0 \\
   \end{array}
 \right)$.
Since
$$\widetilde{A}_\mu = \left(
                       \begin{array}{ccc}
                         0 & 0 & A \\
                         0 & 0 & \mu I \\
                         A^\dag & \mu I & 0 \\
                       \end{array}
                     \right) = \left(
                       \begin{array}{ccc}
                         0 & 0 & A \\
                         0 & 0 & 0 \\
                         A^\dag & 0 & 0 \\
                       \end{array}
                     \right)+\left(
                       \begin{array}{ccc}
                         0 & 0 & 0 \\
                         0 & 0 & \mu I \\
                         0 & \mu I & 0 \\
                       \end{array}
                     \right), $$
The Hamiltonian simulation of $e^{-\i t\widetilde{A}_{\mu}}$ is equivalently efficient as $\left(
   \begin{array}{cc}
     0 & A \\
     A^\dag & 0 \\
   \end{array}
 \right)$
in HHL algorithm \cite{childs17-HS}.
So the assumption about Hamiltonian simulation to solve (\ref{linear-system-regu}) is the same as 
the HHL algorithm. In the following, we always assume that the Hamiltonian simulation of $A_\mu$ is efficient.

Assume that the SVD of $\widetilde{A}_\mu=\sum_j \tilde{\sigma}_j|\tilde{u}_j\rangle\langle \tilde{u}_j|$,
where $\tilde{\sigma}_j$ is the singular value of $\widetilde{A}_\mu$, and the smallest nonzero one is represented by $\tilde{\sigma}_{\min}$.
We formally rewrite $|\tilde{b}\rangle=\sum_j \tilde{\beta}_j |\tilde{u}_j\rangle$.
Then using a similar procedure to the HHL algorithm, we can obtain
\be \label{solving-1}
|\Phi\rangle=\sum_j \tilde{\beta}_j |\tilde{u}_j\rangle
\left[  \widetilde{C}\tilde{\sigma}_j^{-1}|0\rangle + \sqrt{1- \widetilde{C}^2\tilde{\sigma}_j^{-2}} |1\rangle   \right]
= \widetilde{C} \|x_\mu\| |x_\mu\rangle |0\rangle + P_1 |\phi_1\rangle|1\rangle ,
\ee
where $\widetilde{C}= \tilde{\sigma}_{\min}$, and $|\phi_1\rangle$ is proportional to
$\sum_j \tilde{\beta}_j\sqrt{1- \widetilde{C}^2\tilde{\sigma}_j^{-2}} |\tilde{u}_j\rangle$
with amplitude $P_1$.
The complexity to get this quantum state is $O(\kappa_\mu(\log n)/\epsilon)$.
The derivation of this complexity is the same as the HHL algorithm.
However, HHL algorithm has a quadratic dependence on the condition number,
which arises from two resources. One is for estimating $\tilde{\sigma}_j^{-1}$ and the other one is from the success probability.
Here, we do not perform any measurement, so the complexity is linear in the condition number.
We will see in the next section that the quantum state (\ref{solving-1}) is suffices to
find the best regularization parameter.

%

\section{Choice of regularization parameter}

A key issue for the success of the Tikhonov
regularization is how to determine a reasonable regularization
parameter $\mu$. Here we consider the quantum implementations of two typical heuristic methods,
which use the L-curve and the GCV function respectively.
In L-curve or GCV function, we need to estimate the norm of the solution $\|x_\mu\|$
and the norm of the residual $\|Ax_\mu-b\|$.

In (\ref{solving-1}), by amplitude estimation (i.e., Proposition \ref{prop-AA}),
we can estimate the amplitude of $|0\rangle$, which equals $\widetilde{C}\|x_\mu\|$.
By Proposition \ref{prop-AA}, we can obtain an $\alpha$ such that $|\widetilde{C}\|x_\mu\|-\alpha|\leq \epsilon_0$,
that is, $|\|x_\mu\|-\alpha/\widetilde{C}|\leq \epsilon_0/\widetilde{C}$.
This costs $O(\kappa_\mu(\log n)/\epsilon\epsilon_0)$. To make the error small in size $\epsilon$,
we choose $\epsilon_0=\widetilde{C}\epsilon$. Finally, the complexity to get an $\epsilon$-approximation of $\|x_\mu\|$ is
\be \label{cost1}
O(\kappa_\mu(\log n)/\epsilon^2 \tilde{\sigma}_{\min} ).
\ee

Next, we consider how to estimate the norm $\|Ax_\mu-b\|$.
As a generalization of HHL algorithm, the quantum state $|x_\mu\rangle$ given in \eqref{solving-1} can be multiplied by $A$.
And we will get the following quantum state
\be \label{solving-2}
|\psi\rangle =   C \|x_\mu\| A|x_\mu\rangle |0\rangle + P_2 |\phi_2\rangle|1\rangle  ,
\ee
where $C= \widetilde{C}/ \sigma_{\max}$, and $|\phi_2\rangle$ is some garbage state with amplitude $P_2$.
To estimate $\|Ax_\mu-b\|$ by amplitude estimation, we should prepare the quantum state proportional to
$\|x_\mu\|A|x_\mu\rangle-|b\rangle$. It can be obtained in the following steps.

{\bf Step 1}, prepare the initial state as
\be \label{solving-2:eq1}
\frac{1}{\sqrt{2}} \left( |\psi\rangle|0\rangle -  |b,0\rangle |1\rangle \right) |0\rangle.
\ee

{\bf Step 2}, set $t=\min\{1,C\}$. Here we do not need $t$ equals $\min\{1,C\}$ exactly, a low bound of $\min\{1,C\}$ still works.
Apply a control transformation to (\ref{solving-2:eq1}) such that,
if the second register is $|0\rangle$, then change the last qubit $|0\rangle$ into
$
tC^{-1} |0\rangle + \sqrt{1- t^2 C^{-1}} |1\rangle
$.
If the second register  is $|1\rangle$, then change the last qubit $|0\rangle$ into
$
t |0\rangle + \sqrt{1- t^2 } |1\rangle
$.
So we get
\be\label{solving-2:eq2}
\frac{1}{\sqrt{2}} |\psi\rangle |0\rangle \Big[ tC^{-1} |0\rangle + \sqrt{1- t^2 C^{-1}} |1\rangle \Big]
- \frac{1}{\sqrt{2}}  |b,0\rangle |1\rangle \Big[t |0\rangle + \sqrt{1- t^2 } |1\rangle\Big].
\ee

{\bf Step 3}, apply the Hadamard gate to the second register of (\ref{solving-2:eq2}) to get
\be\ba{lll} \label{solving-3} \vspace{.2cm}
&& \ds \frac{1}{\sqrt{2}} |\psi\rangle |+\rangle \Big[ tC^{-1} |0\rangle + \sqrt{1- t^2 C^{-1}} |1\rangle \Big]
- \frac{1}{\sqrt{2}} |b,0\rangle |-\rangle \Big[t |0\rangle + \sqrt{1- t^2 } |1\rangle\Big] \\
&=& \ds \frac{t}{2} (\|x_\mu\| A|x_\mu\rangle - |b\rangle) |0,0,0\rangle + {\rm orthogonal~parts}.
\ea\ee
To get the above quantum state, it only costs $O(\kappa_\mu(\log n)/\epsilon)$ that comes from step 1.
By Proposition \ref{prop-AA}, we will get a value $\beta$ in time $O(\kappa_\mu(\log n)/\epsilon\epsilon_1)$, such that
$
\left| \frac{t}{2} \|\|x_\mu\| A|x_\mu\rangle-|b\rangle\|-\beta\right|\leq \epsilon_1
$.
Similarly, to get an $\epsilon$-approximate of  $\|\|x_\mu\| A|x_\mu\rangle-|b\rangle\|$,
we choose $\epsilon_1 = {\epsilon t}/{ 2 }$. By the definition of $t$ and $C$,
we have $\epsilon_1 \geq \epsilon C/2 = \epsilon \tilde{\sigma}_{\min}/2\sigma_{\max} $.
Finally, the complexity to get an $\epsilon$-approximation of $\|\|x_\mu\| A|x_\mu\rangle-|b\rangle\|$ reads
\be \label{cost2}
O(\kappa_\mu(\log n) \sigma_{\max} /\epsilon^2 \tilde{\sigma}_{\min} ).
\ee

Note that $\tilde{\sigma}_j=\sqrt{\sigma_j^2+\mu^2}$ and generally $\mu \ll 1$.
Since the singular values of $A$ contain small magnitude,
we just assume that $1/\kappa\leq \sigma_j<1$ for all $j$.
Thus,  $\max\tilde{\sigma}_j=O(1)$, which implies $1/ \tilde{\sigma}_{\min}   \approx \kappa_\mu$.
Concluding the above analysis,  we have

\bt \label{thm1}
Assume that the Hamiltonian simulation of
$\left(
   \begin{array}{cc}
     0 & A \\
     A^\dag & 0 \\
   \end{array}
 \right)$ is efficient,
$x_\mu$ is the solution of the regularized least squares system (\ref{x_regu}).
Then we can compute the $\epsilon$-approximations of $\|x_\mu\|$ and $\|Ax_\mu-b\|$ in time
$O(\kappa_\mu^2(\log n)/\epsilon^2)$, where $\kappa_\mu$ is the condition number of
$\left(
\begin{array}{c}
A \\
\mu I \\
\end{array}
\right)$.
\et

\subsection{L-Curve}

The so-called L-curve method uses a plot of the norm of the regularized solution versus the corresponding
residual norm, i.e.,  the plot of $(   \|A x_\mu - b \|,   \| x_\mu\| )$ over a range of $\mu$.
It gives an insight into the regularizing properties of the underlying regularization method, and helps to choose an appropriate regularization parameter.
To be precisely, we set a series of parameters, for example, $\mu_j = \rho^j (j=1,\ldots, p), \rho=0.9$, and plot $  \|A x_\mu - b \|$ vs. $  \| x_\mu\|$.
If there is a corner on the L-curve, one can take the
corresponding parameter $\mu$ as the desired regularization parameter.

Computing all the values of $(  \|A x_{\mu_j} - b \|,   \| x_{\mu_j}\|)$ for $j=1,\ldots, p$,
will costs $O(p(\max_j\kappa_{\mu_j})^2(\log n)/\epsilon^2)$ by Theorem \ref{thm1}.
We use these $p$ pairs of data to plot the L-curve and locate its corner.
Compared with classical method, it achieves an exponential speedup at $n$.

By quantum minimum finding algorithm, i.e., Corollary \ref{coro-AA-appliation}, we can further achieve a quadratic speedup at $p$.
Now we consider the case that $p$ is large. In this case, the appropriate regularization parameter is one of $\mu_j$,
and we can find the best regularization parameter by quantum minimum finding algorithm
and computing $(  \|A x_{\mu_j} - b \|,   \| x_{\mu_j}\|)$ in parallel.
This is achieved in the following steps.

{\bf Step 1}, prepare the state
$$\frac{1}{\sqrt{p}} \sum_{j=1}^p|j\rangle .$$

{\bf Step 2}, denote the quantum state \eqref{solving-1} and \eqref{solving-3} for $\mu_j$ as $|\Phi_j\rangle$
and $|\Psi_j\rangle$ respectively. Then prepare them in parallel by control operation, so we have
$$ \frac{1}{\sqrt{p}} \sum_{j=1}^p|j\rangle |\Phi_j\rangle |\Psi_j\rangle. $$

{\bf Step 3}, apply (\ref{AA:parallel procedure}) to estimate $\|x_{\mu_j}\|$ and $\|A x_{\mu_j} - b \|$ in parallel, which yields
\be \label{state:L-curve}
\frac{1}{\sqrt{p}} \sum_{j=1}^p|j\rangle |\Phi_j\rangle |\Psi_j\rangle |\|x_{\mu_j}\|,\|A x_{\mu_j} - b \|\rangle .
\ee
The function used to compute $\|x_{\mu_j}\|$ from $|\Phi_j\rangle$ is $f(x)=x/\widetilde{C}$,
and the function  used to compute $\|A x_{\mu_j} - b \|$ from $|\Psi_j\rangle$ is $f(x)=2x/t$.
Both are linear functions, so the circuits to  implement their oracles are efficient.

{\bf Step 4}, apply quantum minimum finding algorithm  to find the best $\mu_j$.
In L-curve, the corner corresponds to the optimal regularization parameter.
For most cases, the point $( \|A x_{\mu_j} - b \|, \|x_{\mu_j}\|)$ with minimal $ \|x_{\mu_j}\|^2 +  \|A x_{\mu_j} - b \|^2$
is the corner. Otherwise, a suitable translation is needed.
With the quantum state (\ref{state:L-curve}), we can apply the quantum minimum finding algorithm
to find the minimum of $\{ \|x_{\mu_j}\|^2 +  \|A x_{\mu_j} - b \|^2 :j=1,\ldots,p\}$ and so find the
best $\mu_j$.

It is obvious that step 1 only costs $O(\log p)$. From the construction of \eqref{solving-1} and \eqref{solving-3}, step 2 costs
$O((\max_j\kappa_{\mu_j})(\log n)/\epsilon)$.
In step 3, the amplitude estimation is accomplished in parallel,
so by Corollary \ref{coro-AA} and Theorem \ref{thm1}, this step costs $O((\max_j\kappa_{\mu_j})^2(\log n)/\epsilon^2)$.
Finally, in step 4, apply quantum minimum finding algorithm to find the best $\mu_j$ can achieve a quadratic speedup at $p$.
So the complexity of the above procedure is $O(\sqrt{p}(\max_j\kappa_{\mu_j})^2(\log np)/\epsilon^2)$.


\subsection{GCV function}

The generalized cross-validation (GCV) function
\cite{GolubHeathWahba_Technometrics79} is a choice rule that determines the
regularization parameter by minimizing the GCV function
\begin{equation} \label{gcv}
 G (\mu) = \frac{\|(I - A (A^\dag A + \mu^2 I)^{-1} A^\dag )b\|^2}
{[\text{Tr}(I-A (A^\dag A + \mu^2 I)^{-1} A^\dag  )]^2} 
= \frac{\|A x_\mu - b\|^2}{ [m-n + \sum_{i=1}^n{\mu^2}/({\sigma_i^2 + \mu^2})]^2 }.
\end{equation}
Precisely speaking, we choose a series of regularization parameters $\mu_j~(j=1,\cdots, p)$. For each parameter $\mu_j$, we solve the LSP \eqref{x_regu} and compute the value $G(\mu_j)$. We then use these data to fit a function $G(\mu)$ and seek the optimal parameter that minimizes the function, i.e., $\arg \min_\mu G(\mu)$.
Different from L-curve, another problem that we need to solve in computing the value of
GCV function is the summation $g(\mu) = \mu^2 \sum_{i=1}^n({\sigma_i^2 + \mu^2})^{-1}$.
Calculating such a summation  is not easy in a quantum computer since $\sigma_i$ is unknown, and each measurement only returns one singular value of $A$.
For most of practical problems, $A$ is of low rank, so we can consider an approximate SVD to calculate the GCV function \cite{XiangZou_InverseProb13}. 
That is, we can use a rank-$r$ approximation of $A$ and compute a partial summation $\sum_{i=1}^r({\sigma_i^2 + \mu^2})^{-1}$
with the first $r$ largest singular values.

Let $\widetilde{A}_\mu=(\tilde{a}_{ij})$.
By SVD and the low rank assumption, we have
\[\ba{llll} \vspace{.2cm}
|\widetilde{A}_\mu\rangle &=& \ds \frac{1}{\|\widetilde{A}_\mu\|_F} \sum_{i,j} \tilde{a}_{ij} |i,j\rangle  & {\rm by~definition} \\ \vspace{.2cm}
&=& \ds \frac{1}{\|\widetilde{A}_\mu\|_F} \sum_{i=1}^r \tilde{\sigma}_i |\tilde{u}_i,\tilde{u}_i\rangle
+\frac{1}{\|\widetilde{A}_\mu\|_F} \sum_{i=r+1}^n \tilde{\sigma}_i |\tilde{u}_i,\tilde{u}_i\rangle & {\rm by~SVD} \\
&\approx& \ds \frac{1}{\|\widetilde{A}_\mu\|_F} \sum_{i=1}^r \tilde{\sigma}_i |\tilde{u}_i,\tilde{u}_i\rangle & {\rm by~low~rank~assumption}
\ea\]
Performing the QPE on $\exp(-\i \widetilde{A}_\mu)$ with the initial state $|\widetilde{A}_\mu\rangle$, we will obtain
\be \label{gcv:eq1}
\frac{1}{\|\widetilde{A}_\mu\|_F} \sum_{i=1}^r \tilde{\sigma}_i |\tilde{u}_i,\tilde{u}_i\rangle|\tilde{\sigma}_i\rangle
\ee
with high probability close to 1. Performing $O(r)$  measurements to (\ref{gcv:eq1}), we will obtain all the principal singular values
$\tilde{\sigma}_1,\ldots,\tilde{\sigma}_r$. Equivalently we get the approximations of $\sigma_1,\ldots,\sigma_r$, since the singular values of $\widetilde{A}_\mu$ satisfy $\tilde{\sigma}_i^2 = \sigma_i^2 + \mu^2$.
The complexity of this procedure is $O(r(\log n)/\epsilon)$.
For simplicity, define $g(\mu) = \sum_{i=1}^r {\mu^2}/({\sigma_i^2 + \mu^2})$.
Now, the evaluation of $G(\mu_j)$ reduces to the evaluation of $\|A x_{\mu_j} - b\|^2$
and $g(\mu_j)$. The former can be obtained by \eqref{AA:parallel procedure} and (\ref{solving-3}),  and the latter can be achieved by an oracle to query $\mu_j$.
The following procedure to find the best regularization parameter is similar to that of L-curve.

{\bf Step 1}, prepare the initial state as
\[
\frac{1}{\sqrt{p}} \sum_{j=1}^p|j\rangle|\mu_j\rangle |\Psi_j\rangle,
\]
where $|\Psi_j\rangle$ is the same as that in L-curve.

{\bf Step 2}, apply (\ref{AA:parallel procedure}) to estimate $\|A x_{\mu_j} - b\|^2$ in parallel
and an oracle to calculate $g(\mu_j)$. So we get
\[
\frac{1}{\sqrt{p}} \sum_{j=1}^p|j\rangle|\mu_j\rangle |g(\mu_j)\rangle |\Psi_j\rangle | \|A x_{\mu_j} - b\|^2 \rangle.
\]

{\bf Step 3}, apply an oracle to compute $G(\mu_j) = \frac{\|A x_{\mu_j} - b\|^2}{(m-n+g(\mu_j))^2}$ and store it in an ancilla qubit
\[
\frac{1}{\sqrt{p}} \sum_{j=1}^p|j\rangle|\mu_j\rangle |g(\mu_j)\rangle |\Psi_j\rangle | \|A x_{\mu_j} - b\|^2 \rangle|G(\mu_j)\rangle.
\]

{\bf Step 4}, apply quantum minimum finding algorithm to find the (best) regularization parameter $\mu_{j_0}$ with  $j_0 = \arg \min_j G(\mu_j)$.

By quantum superposition and the procedure to construct $|\Psi_j\rangle$,
step 1 costs $O((\max_j\kappa_{\mu_j})(\log np)/\epsilon)$.
In step 2, the amplitude estimation is accomplished in parallel,
so by Corollary \ref{coro-AA} and Theorem \ref{thm1}, this step costs $O((\max_j\kappa_{\mu_j})^2(\log np)/\epsilon^2)$.
Step 3 is achieved by an efficient oracle.
Finally, in step 4, the quantum minimum finding algorithm costs $O(\sqrt{p}(\max_j\kappa_{\mu_j})^2(\log np)/\epsilon^2)$
to find the best regularization parameter.
Together with the complexity to calculate all principal singular values, the whole complexity of the quantum algorithm to
find the best regularization parameter based on GCV is
$O(r(\log n)/\epsilon+\sqrt{p}(\max_j\kappa_{\mu_j})^2(\log np)/\epsilon^2)$.
For low rank linear system, that is $r=O({\rm poly} \log n)$, then this result is the same as L-curve.
In conclusion, we have

\bt \label{thm2}
Given $p$ regularization parameters $\mu_1,\ldots,\mu_p$, we can find the best regularization parameter in time
$O(\sqrt{p}(\max_j\kappa_{\mu_j})^2(\log np)/\epsilon^2)$ in quantum computer,
where $\kappa_{\mu_j}$ is the condition number of
$\left(
\begin{array}{c}
A \\
\mu_j I \\
\end{array}
\right)$.
\et

\section{Conclusion}

The LSP solver is a basic engine in big data and machine learning. To obtain a meaningful solution for an ill-posed problem, a regularization technique is necessary. The determination of regularization parameter is the most important, but also the most time consuming.
In this paper, based on L-curve and GCV function, we proposed two quantum algorithms to solve the
regularization parameter estimate problem. The result shows that quantum computer can achieve
a quadratic speedup in the number of given regularization parameters and an exponential speedup
in the dimension of problem size.
%
%
The complexity to find the best regularization parameter by L-curve or GCV function depends
on the condition number $\kappa_\mu$ of $A_\mu$.
For an ill-posed problem, the condition number
$\kappa$ of $A$ is often very large. For a properly chosen parameter $\mu$, the condition number $\kappa_\mu$ can be much smaller. But if the regularization parameter $\mu$ is not good, then $\kappa_\mu$ can be still very large, and the HHL solver runs very slowly.
For practical implementation, it is reasonable to set a threshold $\tau = O({\rm poly} \log n)$ for the runtime of HHL.
If the runtime is larger than $\tau$, then we can conclude that the parameter is not good and stop the algorithm.
In this paper,
we only focused on the determination of Tikhonov regularization parameter, but our analysis is also applicable to TSVD.

\end{document}